\documentclass[journal]{IEEEtran}

\usepackage{color}
\usepackage[cmex10]{amsmath}
\usepackage{amssymb}
\usepackage{epsfig}
\usepackage{subfigure}
\usepackage{stfloats}
\usepackage{cite}
\usepackage{enumerate}
\usepackage{graphicx}
\usepackage{amsmath}
\usepackage{bm}
\usepackage{amsmath}
\usepackage{mathtools}
\usepackage{multicol}
\usepackage{yhmath}
\usepackage{polynom}
\usepackage{mathabx}
\usepackage{booktabs}
 \usepackage{float}

\makeatletter
\renewcommand\normalsize{%
   \@setfontsize\normalsize\@xpt\@xiipt
   \abovedisplayskip 5\p@ \@plus2\p@ \@minus5\p@
   \abovedisplayshortskip \z@ \@plus3\p@
   \belowdisplayshortskip 5\p@ \@plus3\p@ \@minus3\p@
   \belowdisplayskip \abovedisplayskip
   \let\@listi\@listI}
\makeatother

\begin{document}
\title{\LARGE Robust and Secure Communications in Intelligent Reflecting\\ Surface Assisted NOMA networks}
\author{Zheng Zhang, Lu Lv, Qingqing Wu, Hao Deng, and Jian Chen
\thanks{Z. Zhang, L. Lv, and J. Chen are with the State Key Laboratory of Integrated Services Networks, Xidian University, Xi'an 710071, China (e-mail: zzhang\_688@stu.xidian.edu.cn; lulv@xidian.edu.cn; jianchen@mail.xidian.edu.cn).}
\thanks{Q. Wu is with the State Key Laboratory of Internet of Things for Smart City, University of Macau, Macau 999078, China (e-mail: qingqingwu@um.edu.mo).}
\thanks{H. Deng is with the School of Physics and Electronics, Henan University, Kaifeng 475001, China (e-mail: gavind@163.com).}
}
\maketitle
\begin{abstract}
This letter investigates secure transmission in an intelligent reflecting surface (IRS) assisted non-orthogonal multiple access (NOMA) network. Consider a practical eavesdropping scenario with imperfect channel state information of the eavesdropper, we propose a robust beamforming scheme using artificial noise to guarantee secure NOMA transmission with the IRS. A joint transmit beamforming and IRS phase shift optimization problem is formulated to minimize the transmit power. Since  the problem is non-convex and challenging to resolve, we develop an effective alternative optimization (AO) algorithm to obtain stationary point solutions. Simulation results validate the security advantage of the robust beamforming scheme and the effectiveness of the AO algorithm.
\end{abstract}

\IEEEpeerreviewmaketitle

\section{Introduction}\label{1:int}
Intelligent reflecting surface (IRS) is a promising technology to achieve high energy and spectrum efficiency for future wireless communications \cite{R.Zhang_IRS_magazine,C.Huang_TWC2019}. Particularly, IRS can actively create a reconfigurable radio environment to improve the wireless network performance by adaptively adjusting amplitudes and phase shifts of passive reflecting elements \cite{Q.Wu_TWC2019}. On the other hand, non-orthogonal multiple access (NOMA) improves spectrum efficiency by exploiting the power-domain multiplexing to serve multiple users with the same time-frequency resource block \cite{L.Lv_magazine,L.Lv_NOMA_PLS}. It is expected that combining IRS with NOMA could further enhance the network performance, since NOMA is more powerful when the differences of the user channel gains are larger, while IRS can proactively reconfigure user channels to achieve this goal \cite{B.Zheng_IRS_NOMA,X.Mu_IRS_NOMA,Z.Ding_IRS_NOMA_1}.

With the broadcast nature of wireless channels, the private information is vulnerable to eavesdropping. This thus calls for physical layer security (PLS), which utilizes the characteristics of wireless channels to achieve secure communications. Since IRS can smartly change the wireless propagation environment, it can be exploited to benefit PLS by intelligently adjusting the reflection coefficients for signal enhancement at receiver while signal cancellation/mitigation at eavesdropper \cite{X.Guan_IRS_AN,H.Shen_IRS_Security,Z.Chu_IRS_Security,X.Yu_imperfectCSI}.

The aforementioned works \cite{X.Guan_IRS_AN,H.Shen_IRS_Security,Z.Chu_IRS_Security,X.Yu_imperfectCSI} only consider PLS for IRS assisted orthogonal multiple access (OMA) networks, while research on PLS for IRS assisted NOMA networks is still missing in the literature. For IRS assisted NOMA with security considerations, resource allocation becomes rather challenging, because: 1) successive interference cancellation (SIC) decoding constraint of NOMA increases the design complexity of the transmission scheme, and 2) the existence of co-channel interference and secrecy constraints lead to sophisticated interference management for IRS's reflection. Furthermore, existing studies \cite{X.Guan_IRS_AN,H.Shen_IRS_Security,Z.Chu_IRS_Security} rely on perfect channel state information (CSI) of eavesdropper which, however, may not hold since eavesdropper is passive and may try to hide itself from legitimate nodes \cite{X.Yu_imperfectCSI}. In this case, only imperfect CSI of eavesdropper is available.

Motivated by the above observations, this letter studies secure transmission in an IRS assisted NOMA network with only the imperfect CSI of a multi-antenna eavesdropper. The major contributions are summarized as follows.
\begin{itemize}
\item We propose a robust beamforming scheme to secure IRS assisted NOMA transmission, where artificial noise (AN) is exploited to reduce information leakage to eavesdropper while minimizing the effect on reception quality of legitimate users. A joint active and passive beamforming optimization problem is formulated and solved for transmit power minimization.
\item To handle the non-convex constraints due to the eavesdropper's imperfect CSI, we introduce equivalent channel/beamforming matrices to simplify the semi-infinite constraints. Furthermore, a sequential rank-one constraint relaxation (SROCR) based alternative optimization (AO) algorithm is proposed to efficiently optimize the IRS reflection coefficients and the transmit power, where effective rank-one solutions are obtained.
\item Numerical results verify the security advantage of proposed scheme over two baseline schemes. In particular, it is found that signal power highly depends on quality-of-service (QoS) constraint of legitimate users, while AN power is extremely sensitive to QoS constraint, maximum eavesdropping rate, and interception capability of eavesdropper.
\end{itemize}

\section{System Model and Problem Formulation}\label{2:sys}

\begin{figure}[t]
\centering
\includegraphics[scale=0.3]{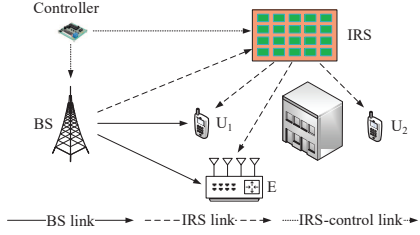}
\caption{An IRS assisted NOMA network.}
\label{fig1:model}
\end{figure}

As depicted in Fig. \ref{fig1:model}, we consider an IRS assisted NOMA network, which consists of two single-antenna users ($\text{U}_{1}$ and $\text{U}_{2}$), an $N_{\text{t}}$-antenna base station (BS), an IRS, and an $N_{\text{e}}$-antenna eavesdropper (E). We assume that there is no direct link between the BS and $\text{U}_{2}$ due to the existence of obstacles, which thus needs the deployment of IRS to establish a reliable communication link between them. As $\text{U}_{1}$ and E locate close to the BS, they have direct links to the BS. In each transmission, the BS utilizes NOMA to simultaneously transmit superimposed signals and AN, where AN is used to confuse E. The IRS is connected to a smart controller and has $M$ passive reflecting elements, each of which can change its amplitude and phase independently to improve the reception quality of both $\text{U}_{1}$ and $\text{U}_{2}$ while degrading the interception capability of E.

The received signals at $\text{U}_{1}$, $\text{U}_{2}$ and E are given by
\begin{equation}
\label{1}
y_{1}=(\mathbf{h}_{\text{I},1}^{H}\bm{\Theta}\mathbf{H}_{\text{B},\text{I}}+\mathbf{h}_{\text{B},1}^{H})(\sum\nolimits_{i=1}^{2}\mathbf{w}_{i}s_{i}
+\mathbf{s}_{\text{AN}})+n_{1},
\end{equation}
\begin{equation}
\label{2}
y_{2}=(\mathbf{h}_{\text{I},2}^{H}\bm{\Theta}\mathbf{H}_{\text{B},\text{I}})(\sum\nolimits_{i=1}^{2}\mathbf{w}_{i}s_{i}+\mathbf{s}_{\text{AN}})+n_{2},
\end{equation}
\begin{equation}
\label{3}
\mathbf{y}_{\text{e}}=(\mathbf{G}_{\text{I},\text{e}}^{H}\bm{\Theta}\mathbf{H}_{\text{B},\text{I}}+\mathbf{G}_{\text{B},\text{e}}^{H})(\sum\nolimits_{i=1}^{2}
\mathbf{w}_{i}s_{i}+\mathbf{s}_{\text{AN}})+\mathbf{n}_{\text{e}},
\end{equation}
where $\mathbf{h}_{\text{I},i}\in\mathbb{C}^{M\times 1}$ ($1\leq i\leq 2$), $\mathbf{h}_{\text{B},1}\in\mathbb{C}^{N_{\text{t}}\times 1}$, $\mathbf{H}_{\text{B},\text{I}}\in\mathbb{C}^{M\times N_{\text{t}}}$ $\mathbf{G}_{\text{B},\text{e}}\in\mathbb{C}^{N_{\text{t}}\times N_{\text{e}}}$, and $\mathbf{G}_{\text{I},\text{e}}\in\mathbb{C}^{M\times N_{\text{e}}}$ denote the channel vectors/matrixes of transmission links IRS-$\text{U}_{i}$ ($1\leq i\leq 2$), BS-$\text{U}_{1}$, BS-IRS, BS-E, and IRS-E, respectively. $s_{i}$ denotes the signal of $\text{U}_{i}$ with the corresponding beamfoming vector $\mathbf{w}_{i}\in\mathbb{C}^{N_{\text{t}}\times 1}$, which satisfies $\mathbb{E}\{|s_{i}|^{2}\}=1$. $\mathbf{s}_{\text{AN}}\in\mathbb{C}^{N_{\text{t}}\times 1}$ denotes the AN vector following circularly symmetric complex Gaussian distribution with zero mean and covariance matrix $\mathbf{W}_{\text{AN}}$. $n_{1},n_{2}\sim\mathcal{CN}(0,1)$ and $\mathbf{n}_{\text{e}}\sim\mathcal{CN}(\mathbf{0},\mathbf{I}_{N_{\text{e}}})$ are the additive white Gaussian noises (AWGNs) at users and E, respectively. To explore the fundamental performance limit of the considered network, we assume that the reflection coefficients of the IRS can be arbitrary amplitudes and phase shift values, i.e., $\bm{\Theta}=\text{diag}(\beta_{1}e^{j\theta_{1}},\dots,\beta_{M}e^{j\theta_{M}})\in\mathbb{C}^{M\times M}$, where $\beta_{m}\in[0,1]$ and $\theta_{m}\in[0,2\pi]$ for $1\leq m\leq M$.

Without loss of generality, the channel gains are ordered as $\|\mathbf{h}_{1}\|^{2}\geq \|\mathbf{h}_{2}\|^{2}$, where $\mathbf{h}_{1}^{H}=\mathbf{h}_{\text{I},1}^{H}\bm{\Theta}\mathbf{H}_{\text{B},\text{I}}+\mathbf{h}_{\text{B},1}^{H}$ and $\mathbf{h}_{2}^{H}=\mathbf{h}_{\text{I},2}^{H}\bm{\Theta}\mathbf{H}_{\text{B},\text{I}}$. As for $\text{U}_{2}$'s signal decoding, $s_{2}$ is detected firstly by treating $s_{1}$ as noise at each receiver, then $\text{U}_{1}$
will remove $s_{2}$ from the detect result and decoding $s_{1}$ without inter-user interference. Accordingly, the achievable rates for $\text{U}_{1}$ to decode $s_{1}$ and $s_{2}$ are given, respectively, by
\begin{equation}
\label{4}
R_{1,1}= \log_{2}\left(1+\frac{|\mathbf{h}_{1}^{H}\mathbf{w}_{1}|^{2}}
{\text{Tr}(\mathbf{h}_{1}^{H}\mathbf{h}_{1}\mathbf{W}_{\text{AN}})+1}\right),
\end{equation}
\begin{equation}
\label{5}
R_{1,2}= \log_{2}\left(1+\frac{|\mathbf{h}_{1}^{H}\mathbf{w}_{2}|^{2}}
{|\mathbf{h}_{1}^{H}\mathbf{w}_{1}|^{2}+\text{Tr}(\mathbf{h}_{1}^{H}\mathbf{h}_{1}\mathbf{W}_{\text{AN}})+1}\right).
\end{equation}
While $\text{U}_{2}$ directly decodes $s_{2}$ by treating $s_{1}$ as noise yielding the achievable rate as
\begin{equation}
\label{6}
R_{2,2}= \log_{2}\left(1+\frac{|\mathbf{h}_{2}^{H}\mathbf{w}_{2}|^{2}}
{|\mathbf{h}_{2}^{H}\mathbf{w}_{1}|^{2}+\text{Tr}(\mathbf{h}_{2}^{H}\mathbf{h}_{2}\mathbf{W}_{\text{AN}})+1}\right).
\end{equation}
We adopt a worst-case assumption in PLS, namely, E has strong multiuser detection capacity and can remove inter-user interference in NOMA secrecy \cite{L.Lv_Secure_MISO}. Thus, the eavesdropping rates at E for $s_{1}$ and $s_{2}$ are shown as
\begin{equation}
\label{7}
R_{\text{e},1}= \log_{2}\text{det}(\mathbf{I}_{N_{\text{e}}}+\mathbf{Q}^{-1}
\mathbf{G}_{\text{e}}^{H}
\mathbf{w}_{1}\mathbf{w}_{1}^{H}\mathbf{G}_{\text{e}}),
\end{equation}
\begin{equation}
\label{8}
R_{\text{e},2}= \log_{2}\text{det}(\mathbf{I}_{N_{\text{e}}}+\mathbf{Q}^{-1}
\mathbf{G}_{\text{e}}^{H}
\mathbf{w}_{2}\mathbf{w}_{2}^{H}\mathbf{G}_{\text{e}}),
\end{equation}
where $\mathbf{G}_{\text{e}}^{H}=\mathbf{G}_{\text{I},\text{e}}^{H}\bm{\Theta}\mathbf{H}_{\text{B},\text{I}}+\mathbf{G}_{\text{B},\text{e}}^{H}$ and $\mathbf{Q}=\mathbf{G}_{\text{e}}^{H}\mathbf{W}_{\text{AN}}\mathbf{G}_{\text{e}}+\mathbf{I}_{N_{\text{e}}}$.

\subsection{Imperfect Channel State Information}\label{2-2:CSI}
In this paper, we assume that the CSI of legitimate users is perfectly available to BS, which can be realized by the channel estimation method mentioned in \cite{R.Zhang_IRS_magazine} and its follow-up works, e.g., the semi-passive channel estimation method. While the perfect CSI of E is difficult to obtain since E usually belongs to third party networks and has no cooperation with the BS. In other words, E only exchanges data with its own network nodes. In this case, BS can utilize the pilot information leakage from E to estimate the CSI, which, however, is inexact and outdated \cite{X.Yu_imperfectCSI}. To describe E's imperfect CSI, we adopt the ellipsoidal bounded channel uncertainty model as follows
\begin{equation}
\label{9}
\Delta\mathbf{G}_{\text{I},\text{e}} = \mathbf{G}_{\text{I},\text{e}}-\mathbf{\hat{G}}_{\text{I},\text{e}},\
\Delta\mathbf{G}_{\text{B},\text{e}} = \mathbf{G}_{\text{B},\text{e}}-\mathbf{\hat{G}}_{\text{B},\text{e}},
\end{equation}
\begin{equation}
\label{10}
\Omega_{\text{e}}=\{\|\Delta\mathbf{G}_{\text{I},\text{e}}\|_{F}\leq \varepsilon_{\text{I},\text{e}},\|\Delta\mathbf{G}_{\text{B},\text{e}}\|_{F}\leq \varepsilon_{\text{B},\text{e}}\},
\end{equation}
where $\mathbf{\hat{G}}_{\text{I},\text{e}}$ and $\mathbf{\hat{G}}_{\text{B},\text{e}}$ represent the estimated channels of $\mathbf{G}_{\text{I},\text{e}}$ and $\mathbf{G}_{\text{B},\text{e}}$, while $\varepsilon_{\text{I},\text{e}}>0$ and $\varepsilon_{\text{B},\text{e}}>0$ denote the sizes of the uncertainty regions of channel estimation errors $\Delta\mathbf{G}_{\text{I},\text{e}}$ and $\Delta\mathbf{G}_{\text{B},\text{e}}$, respectively, and $\|\cdot\|_{F}$ is the Frobenius norm.

\subsection{Problem Formulation}\label{2-3:pro}
To guarantee robust and secure transmission, a worst-case optimization problem is investigated. Specifically, we aim at minimizing the total transmit power by joint active and passive beamforming, subject to the minimum QoS constraints at users and the maximum eavesdropping rates at E. The optimization problem is formulated as follows
\begin{subequations}
\begin{align}
\label{11a} &\min\limits_{\mathbf{w}_{i},\mathbf{W}_{\text{AN}},\bm{\Theta}}\quad \sum\nolimits_{i = 1}^{2}\|\mathbf{w}_{i}\|^{2}+\text{Tr}(\mathbf{W}_{\text{AN}})\\
\label{11b}&\quad\text{s.t.} \quad\,\,\, \theta_{m}\in[0,2\pi],\ \beta_{m}\in[0,1],\ \forall m,\\
\label{11c}&\quad\quad\quad\,\, R_{i,i} \geq R_{\text{Q}},\ \forall i,\\
\label{11d}&\quad\quad\quad\,\, \max\limits_{\Omega_{\text{e}}}R_{\text{e},i} \leq R_{\text{M}},\ \forall i,\\
\label{11e}&\quad\quad\quad\,\,  R_{1,2} \geq R_{2,2},\\
\label{11f}&\quad\quad\quad\,\,  \|\mathbf{h}_{1}\|^{2}\geq \|\mathbf{h}_{2}\|^{2},
\end{align}
\end{subequations}
where $R_{\text{Q}}$ denotes the QoS requirement of users and $R_{\text{M}}$ denotes the maximum eavesdropping rate at E.
Constraint \eqref{11b} represents the IRS amplitudes/phase shifts requirements. Constraints \eqref{11c} and \eqref{11d} guarantee a positive rate gap between legitimate transmission rates and eavesdropping rates. The inequality in \eqref{11e} insures a successful SIC decoding at $\text{U}_{1}$. Constraint \eqref{11f} ensures the SIC decoding order of the NOMA users. Problem \eqref{11a} is intractable to solve due to the semi-infinite constraints \eqref{11d} and coupled variables $\bm{\Theta}$ and $\mathbf{w}_{i}$. Next, we develop an efficient AO algorithm to solve it.

\section{Robust Beamforming Design}\label{3:AO-m}

This section proposes an AO algorithm to efficiently solve problem \eqref{11a}. Specifically, we first introduce equivalent channel/beamforming matrices to transform the semi-infinite constraints into a tractable form, which can be directly tackled by S-procedure. Then, to handle the non-convex constraints caused by the coupled variables, we optimize the active and passive beamforming in an alternative manner.
\subsection{Transformation of Semi-Infinite Constraint}\label{3-1:S-pr}

According to \cite[Prop. 1]{X.Yu_imperfectCSI}, we first rewrite \eqref{11d} into the following form:
\begin{align}
\label{12}
\mathbf{G}_{\text{e}}^{H}
\mathbf{W}'_{i}
\mathbf{G}_{\text{e}}+(2^{R_{\text{M}}}-1)\mathbf{I}_{N_{\text{e}}}\succeq \mathbf{0},\ \forall i,
\end{align}
where $\mathbf{W}'_{i} = (2^{R_{\text{M}}}-1)\mathbf{W}_{\text{AN}}-\mathbf{W}_{i}$ and $\mathbf{W}_{i} = \mathbf{w}_{i}\mathbf{w}_{i}^{H}$ for $1\leq i \leq 2$, which should satisfy constraints $\mathbf{W}_{i}\succeq\mathbf{0}$ and $\text{rank}(\mathbf{W}_{i})=1$. Then, substituting \eqref{9} into \eqref{12}, \eqref{11d} can be further expressed as the following quadratic form:
\begin{align}
\label{13}\nonumber
&(\mathbf{\hat{G}}_{\text{I},\text{e}}^{H}\bm{\Theta}\mathbf{H}_{\text{B},\text{I}}+
\mathbf{\hat{G}}_{\text{B},\text{e}}^{H})\mathbf{W}'_{i}(\mathbf{\hat{G}}_{\text{I},\text{e}}^{H}\bm{\Theta}\mathbf{H}_{\text{B},\text{I}}+
\mathbf{\hat{G}}_{\text{B},\text{e}}^{H})^{H}+\\ \nonumber
&(\Delta\mathbf{G}_{\text{I},\text{e}}^{H}\bm{\Theta}\mathbf{H}_{\text{B},\text{I}}+
\Delta\mathbf{G}_{\text{B},\text{e}}^{H})\mathbf{W}'_{i}(\mathbf{\hat{G}}_{\text{I},\text{e}}^{H}\bm{\Theta}\mathbf{H}_{\text{B},\text{I}}+
\mathbf{\hat{G}}_{\text{B},\text{e}}^{H})^{H}+\\ \nonumber
&(\mathbf{\hat{G}}_{\text{I},\text{e}}^{H}\bm{\Theta}\mathbf{H}_{\text{B},\text{I}}+
\mathbf{\hat{G}}_{\text{B},\text{e}}^{H})\mathbf{W}'_{i}(\Delta\mathbf{G}_{\text{I},\text{e}}^{H}\bm{\Theta}\mathbf{H}_{\text{B},\text{I}}+
\Delta\mathbf{G}_{\text{B},\text{e}}^{H})^{H}+\\ \nonumber
&(\Delta\mathbf{G}_{\text{I},\text{e}}^{H}\bm{\Theta}\mathbf{H}_{\text{B},\text{I}}+
\Delta\mathbf{G}_{\text{B},\text{e}}^{H})\mathbf{W}'_{i}(\Delta\mathbf{G}_{\text{I},\text{e}}^{H}\bm{\Theta}\mathbf{H}_{\text{B},\text{I}}+
\Delta\mathbf{G}_{\text{B},\text{e}}^{H})^{H}+\\
&(2^{R_{\text{M}}}-1)\mathbf{I}_{N_{\text{e}}}\succeq \mathbf{0},\ \forall i.
\end{align}
To handle \eqref{13}, we define the equivalent channel estimation error and estimated channel matrices of E as
\begin{equation}
\label{14}
\Delta\mathbf{X}^{H}=[\Delta\mathbf{G}_{\text{I},\text{e}}^{H},\Delta\mathbf{G}_{\text{B},\text{e}}^{H}],\
\mathbf{\hat{X}}^{H}=[\mathbf{\hat{G}}_{\text{I},\text{e}}^{H},\mathbf{\hat{G}}_{\text{B},\text{e}}^{H}].
\end{equation}
Furthermore, a joint beamforming matrix is defined as
\begin{equation}
\label{15}
\mathbf{V}_{i}=\begin{bmatrix}\bm{\Theta}\mathbf{H}_{\text{B},\text{I}}\mathbf{W}'_{i}\mathbf{H}_{\text{B},\text{I}}^{H}\bm{\Theta}^{H}&
\bm{\Theta}\mathbf{H}_{\text{B},\text{I}}\mathbf{W}'_{i}\\
\mathbf{W}'_{i}\mathbf{H}_{\text{B},\text{I}}^{H}\bm{\Theta}^{H}&\mathbf{W}'_{i}
\end{bmatrix},\ \forall i.
\end{equation}
Thus, combining \eqref{13}, \eqref{14} and \eqref{15}, we obtain
\begin{align}
\label{16} \nonumber
&\Delta\mathbf{X}^{H}\mathbf{V}_{i}\Delta\mathbf{X}+\Delta\mathbf{X}^{H}\mathbf{V}_{i}\mathbf{\hat{X}}+\mathbf{\hat{X}}^{H}\mathbf{V}_{i}\Delta\mathbf{X}+
\mathbf{\hat{X}}^{H}\mathbf{V}_{i}\mathbf{\hat{X}}+\\
&(2^{R_{\text{M}}}-1)\mathbf{I}_{N_{\text{e}}}\succeq \mathbf{0},\Delta\mathbf{X}\in\{\mathbf{Y}|\text{Tr}(\varepsilon_{\text{e}}^{-2}\mathbf{Y}\mathbf{Y}^{H})\leq 1\},\ \forall i,
\end{align}
where $\varepsilon_{\text{e}}=\varepsilon_{\text{B},\text{e}}+\varepsilon_{\text{I},\text{e}}$. Afterwards, by adopting S-procedure \cite{Z.-Q.Luo_S-Procedure}, the infinite inequality \eqref{16} can be transformed into a finite LMI as
\begin{equation}
\label{17}
\begin{bmatrix}
\mathbf{\hat{X}}^{H}\mathbf{V}_{i}\mathbf{\hat{X}}+
(\gamma_{\text{M}}-\tau_{i})\mathbf{I}_{N_{\text{e}}} & \mathbf{\hat{X}}^{H}\mathbf{V}_{i} \\
\mathbf{V}_{i}\mathbf{\hat{X}}& \mathbf{V}_{i}+\tau_{i}\varepsilon_{\text{e}}^{-2}\mathbf{I}_{M}
\end{bmatrix} \succeq \mathbf{0},\ \forall i,
\end{equation}
where $\gamma_{\text{M}}=2^{R_{\text{M}}}-1$, and $\tau_{i}>0$ denotes an auxiliary variable introduced by S-procedure. 

\subsection{Active Beamforming Optimization}\label{3-2:Act}
By fixing $\bm{\Theta}$, the optimization problem becomes:
\begin{subequations}
\begin{align}
\label{18a} &\min\limits_{\mathbf{W}_{i},\mathbf{W}_{\text{AN}},\tau_{i}}\quad \sum\nolimits_{i = 1}^{2}\text{Tr}(\mathbf{W}_{i})+\text{Tr}(\mathbf{W}_{\text{AN}})\\
\label{18b}&\quad\quad\text{s.t.} \quad\,\, \eqref{11c},\eqref{11e},\eqref{17},\\
\label{18c}&\quad\quad\quad\quad\,\,\mathbf{W}_{i}\succeq\mathbf{0},\ \forall i,\\
\label{18d}&\quad\quad\quad\quad\,\, \text{rank}(\mathbf{W}_{i})=1,\ \forall i.
\end{align}
\end{subequations}
For notation brevity, we denote $\mathbf{H}_{\text{W},1}=\mathbf{h}_{1}\mathbf{h}_{1}^{H}$ and $\mathbf{H}_{\text{W},2}=\mathbf{h}_{2}\mathbf{h}_{2}^{H}$. Thus, constraint \eqref{11c} can be rewritten as
\begin{subequations}
\begin{equation}
\label{19a}
\text{Tr}(\mathbf{H}_{\text{W},1}\mathbf{W}_{1})\geq \gamma_{\text{Q}}(\text{Tr}(\mathbf{H}_{\text{W},1}\mathbf{W}_{\text{AN}})+1),
\end{equation}
\begin{equation}\label{19b}
\text{Tr}(\mathbf{H}_{\text{W},2}\mathbf{W}_{2})\geq \gamma_{\text{Q}}(\text{Tr}(\mathbf{H}_{\text{W},2}\mathbf{W}_{\text{AN}})+\text{Tr}(\mathbf{H}_{\text{W},2}\mathbf{W}_{1})+1),
\end{equation}
\end{subequations}
where $\gamma_{\text{Q}}=2^{R_{\text{Q}}}-1$. For constraint \eqref{11e}, we introduce a slack variable $\gamma_{t}>0$, which satisfies
\begin{subequations}
\begin{equation}
\label{20a}
\text{Tr}(\mathbf{H}_{\text{W},1}\mathbf{W}_{2})\geq (\text{Tr}(\mathbf{H}_{\text{W},1}\mathbf{W}_{\text{AN}})+\text{Tr}(\mathbf{H}_{\text{W},1}\mathbf{W}_{1})+1)\gamma_{t},
\end{equation}
\begin{equation}
\label{20b}
\text{Tr}(\mathbf{H}_{\text{W},2}\mathbf{W}_{2})\leq (\text{Tr}(\mathbf{H}_{\text{W},2}\mathbf{W}_{\text{AN}})+\text{Tr}(\mathbf{H}_{\text{W},2}\mathbf{W}_{1})+1)\gamma_{t}.
\end{equation}
\end{subequations}
In \eqref{20a}, it is not difficult to see that the term of $\text{Tr}(\mathbf{H}_{\text{W},1}\mathbf{W}_{\text{AN}})+\text{Tr}(\mathbf{H}_{\text{W},1}\mathbf{W}_{1})+1$ is nonnegative. Thus, we apply the arithmetic geometry mean (AGM) inequality to approximate \eqref{20a} by
\begin{align}
\label{21}\nonumber
2\text{Tr}(\mathbf{H}_{\text{W},1}\mathbf{W}_{2})\geq & ((\text{Tr}(\mathbf{H}_{\text{W},1}\mathbf{W}_{\text{AN}})
+\text{Tr}(\mathbf{H}_{\text{W},1}\mathbf{W}_{1})+1)\varpi)^{2}\\
&+\left(\gamma_{t}/\varpi\right)^{2},
\end{align}
where the equality holds if and only if when $\varpi=\sqrt{\frac{\gamma_{t}}{\text{Tr}(\mathbf{H}_{\text{W},1}\mathbf{W}_{\text{AN}})+\text{Tr}(\mathbf{H}_{\text{W},1}
\mathbf{W}_{1})+1}}$.
In \eqref{20b}, we introduce another slack variable $\nu$, which satisfies
\begin{equation}
\label{22}
\text{Tr}(\mathbf{H}_{\text{W},2}\mathbf{W}_{2})\leq 2 \tilde{\nu}\nu-\tilde{\nu}^{2},
\end{equation}
where the right-hand side of \eqref{22} is the Taylor series expansion of the quadratic function $\nu^{2}$, and $\tilde{\nu}$ denotes the reference point of $\nu$. Then, \eqref{20b} can be reshaped as
\begin{equation}
\label{23}
\begin{bmatrix}
\text{Tr}(\mathbf{H}_{\text{W},2}\mathbf{W}_{\text{AN}})+\text{Tr}(\mathbf{H}_{\text{W},2}\mathbf{W}_{1})+1 & \nu \\
\nu & \gamma_{t}
\end{bmatrix}
\succeq \mathbf{0}.
\end{equation}
To deal with with the non-convex rank-one constraints \eqref{18d}, we consider the SROCR method \cite{P.Cao_SROCR} to obtain rank-one solutions of problem \eqref{18a}, which is described as follows. The rank-one constraint $\text{rank}(\mathbf{W}_{i}^{(n)})=1$ at \textit{n}th iteration is replaced by the linear constraint
\begin{equation}
\label{24}
\mathbf{w}_{i}^{\text{eig-max},(n-1)}\mathbf{W}_{i}^{(n)}\mathbf{w}_{i}^{\text{eig-max},(n-1)}\geq w_{i}^{(n-1)}\text{Tr}(\mathbf{W}_{i}^{(n)}).
\end{equation}
In \eqref{24},
$w_{i}^{(n-1)}\in [0,1]$ denotes the trace ratio parameter of $\mathbf{W}_{i}$ at $(\textit{n}-1)$th iteration, which gradually increases from 0 to 1. $\mathbf{w}_{i}^{\text{eig-max},(n-1)}\in\mathbb{C}^{N_{\text{t}}\times 1}$ denotes the eigenvector of the largest eigenvalue of $\mathbf{W}_{i}^{(n-1)}$ with the parameter $w_{i}^{(n-1)}$. The iterative convex program (ICP) at $n$th iteration is given by
\begin{subequations}
\begin{align}
\label{25a} &\min\limits_{\mathbf{W}_{i},\mathbf{W}_{\text{AN}},\tau_{i},\gamma_{t},\nu}\quad \sum\nolimits_{i = 1}^{2}\text{Tr}(\mathbf{W}_{i})+\text{Tr}(\mathbf{W}_{\text{AN}})\\ \nonumber
\label{25b}&\quad\quad\text{s.t.} \quad\,\, \eqref{17},\eqref{18c},\eqref{19a},\eqref{19b},\eqref{21},\\
&\quad\quad\quad\quad\eqref{22},\eqref{23},\eqref{24}.
\end{align}
\end{subequations}
The ICP can be solved efficiently by using the CVX toolbox, and the iterative algorithm for problem \eqref{25a} is summarized in \textbf{Algorithm-1}, where $P_{\text{t}}$ denotes the total transmit power at BS and $\delta$ denotes the convergence accuracy.
\begin{table}[t]
    \centering
    \begin{tabular}{p{235pt}}
    \toprule
    \textbf{Algorithm-1:} Iterative Algorithm for Solving Problem \eqref{18a} \\
    \midrule
    1: \textbf{Initialization}: set $n = 1$ and initialize $\varpi^{(0)}$, $\tilde{\nu}^{(0)}$, $w_{i}^{(0)}$, $\mathbf{w}_{i}^{\text{eig-max},(0)}$;\\
    2: \textbf{Repeat:} \\
    3: \quad \textbf{If} the ICP \eqref{25a} is feasible, solve the problem, define $\epsilon^{(n)}=$\\
     \qquad $\epsilon^{(n-1)}$ and update $\varpi^{(n)}$ and $\tilde{\nu}^{(n)}$;\\
    4: \quad \textbf{Else}: define $\epsilon^{(n)}=\frac{1}{2}\epsilon^{(n-1)}$;\\
    5: \quad Update $w_{i}^{(n)}=\text{min}(1,\frac{\lambda_{\text{max}}(\mathbf{W}_{i}^{(n)})}{\text{Tr}(\mathbf{W}_{i}^{(n)})}+\epsilon^{(n)})$; \\
    6: \quad$n=n+1$;\\
    7: \textbf{Until:} $w_{i}^{(n)}= 1$ and $|P_{\text{t}}^{(n)}-P_{\text{t}}^{(n-1)}|\leq \delta$.\\
    \bottomrule
    \end{tabular}
\end{table}


\subsection{Passive Beamforming Optimization}\label{3-3:Pas}
With the given $\mathbf{W}_{i}$ and $\mathbf{W}_{\text{AN}}$, we can denote $\mathbf{u}_{0}=[\beta_{1}e^{j\theta_{1}},\dots, \beta_{M}e^{j\theta_{M}}]^{H}$, $\mathbf{u}=[\mathbf{u}_{0};1]$, $\mathbf{U}\triangleq\mathbf{u}\mathbf{u}^{H}$, $\mathbf{H}_{\text{U},1} = [\text{diag}(\mathbf{h}_{\text{I},1}^{H})\mathbf{H}_{\text{B},\text{I}};\mathbf{h}_{\text{B},1}^{H}]$ and $\mathbf{H}_{\text{U},2} = [\text{diag}(\mathbf{h}_{\text{I},2}^{H})\mathbf{H}_{\text{B},\text{I}};\mathbf{0}]$. Therefore, the constraint \eqref{19a} and \eqref{19b} can be transformed into
\begin{subequations}
\begin{equation}
\label{26a}
\text{Tr}(\mathbf{U}'_{1,1})\geq \gamma_{\text{Q}}(\text{Tr}(\mathbf{U}'_{\text{AN},1})+1),
\end{equation}
\begin{equation}
\label{26b}
\text{Tr}(\mathbf{U}'_{2,2})\geq \gamma_{\text{Q}}(\text{Tr}(\mathbf{U}'_{\text{AN},2})
+\text{Tr}(\mathbf{U}'_{1,2})+1),
\end{equation}
\end{subequations}
where $\mathbf{U}'_{\varrho,i}=\mathbf{H}_{\text{U},i}\mathbf{W}_{\varrho}\mathbf{H}_{\text{U},i}^{H}\mathbf{U}$ for $\varrho\in\{1,2,\text{AN}\}$ and $i\in\{1,2\}$. Similarly, the constraints \eqref{21}, \eqref{22} and \eqref{23} can be rewritten as
\begin{align}
\label{27}\nonumber
2\text{Tr}(\mathbf{U}'_{2,1})\geq&((\text{Tr}(\mathbf{U}'_{\text{AN},1})+\text{Tr}(\mathbf{U}'_{1,1})+1)\varpi)^{2}\\
&+\left(\gamma_{t}/\varpi\right)^{2},
\end{align}
\begin{equation}
\label{28}
\text{Tr}(\mathbf{U}'_{2,2})\leq 2 \tilde{\nu}\nu-\tilde{\nu}^{2},
\end{equation}
\begin{equation}
\label{29}
\begin{bmatrix}
\text{Tr}(\mathbf{U}'_{\text{AN},2})+\text{Tr}(\mathbf{U}'_{1,2})+1 & \nu \\
\nu & \gamma_{t}
\end{bmatrix}
\succeq \mathbf{0}.
\end{equation}
Furthermore, we denote $\mathbf{Q}_{i}=\text{diag}(\mathbf{h}_{\text{I},i}^{H})\mathbf{H}_{\text{B},\text{I}}$, $\mathbf{J}_{1}=\begin{bmatrix} \mathbf{Q}_{1}\mathbf{Q}_{1}^{H} & \mathbf{Q}_{1}\mathbf{h}_{\text{B},1} \\ \mathbf{h}_{\text{B},1}^{H}\mathbf{Q}_{1}^{H} & 0 \end{bmatrix}$ and $\mathbf{J}_{2}=\begin{bmatrix} \mathbf{Q}_{2} \\ \mathbf{0} \end{bmatrix}$$\begin{bmatrix} \mathbf{Q}_{2}^{H} & \mathbf{0} \end{bmatrix}$. Then, we transform the constraint \eqref{11f} into:
\begin{equation}
\label{30}
\text{Tr}(\mathbf{J}_{1}\mathbf{U})+\|\mathbf{h}_{\text{B},1}^{H}\|^{2}\geq \text{Tr}(\mathbf{J}_{2}\mathbf{U}).
\end{equation}
For the non-convex term $\mathbf{V}_{i}$ in \eqref{15}, we utilize the singular value decomposition (SVD) to transform $\mathbf{H}_{\text{B},\text{I}}\mathbf{W}'_{i}\mathbf{H}_{\text{B},\text{I}}^{H}$ into $\sum_{p}\mathbf{s}_{i,p}\mathbf{d}_{i,p}$, which represents $\bm{\Theta}\mathbf{H}_{\text{B},\text{I}}\mathbf{W}'_{i}\mathbf{H}_{\text{B},\text{I}}^{H}\bm{\Theta}^{H}$ in an equivalent form $\sum_{p}\text{diag}(\mathbf{s}_{i,p})\mathbf{\bar{u}}_{0}\mathbf{\bar{u}}_{0}^{H}\text{diag}(\mathbf{d}_{i,p})$. Based on the properties of matrix, we have the following equation: 
\begin{equation}
\label{31}
\text{diag}(\mathbf{s}_{i,p})\mathbf{\bar{u}}_{0}\mathbf{\bar{u}}_{0}^{H}\text{diag}(\mathbf{d}_{i,p})=\mathbf{S}_{i,p}\mathbf{\bar{u}}\mathbf{\bar{u}}^{H}\mathbf{D}_{i,p},\ \forall i,p,
\end{equation}
where $\mathbf{S}_{i,p}=\begin{bmatrix}\text{diag}(\mathbf{s}_{i,p}),\mathbf{0}\end{bmatrix}$ and $\mathbf{D}_{i,p}=\begin{bmatrix} \text{diag}(\mathbf{d}_{i,p})\\ \mathbf{0}\end{bmatrix}$. Moreover, when the condition of $\text{rank}(\mathbf{U}) = 1$ holds, $\bm{\Theta}$ can be denoted by $\text{diag}(\mathbf{U}_{N+1,1:N})$ equivalently, where $\mathbf{U}_{N+1,1:N}$ equals to  $[\mathbf{U}_{N+1,1},\dots,\mathbf{U}_{N+1,N}]$. Therefore, non-convex term $\mathbf{V}_{i}$ can be rewritten as
$\begin{bmatrix}\sum_{p}\mathbf{S}_{i,p}\mathbf{U}^{T}\mathbf{D}_{i,p}&
\text{diag}(\mathbf{U}_{N+1,1:N})\mathbf{H}_{\text{B},\text{I}}\mathbf{W}'_{i}\\
\mathbf{W}'_{i}\mathbf{H}_{\text{B},\text{I}}^{H}\text{diag}(\mathbf{U}_{N+1,1:N})^{H}&\mathbf{W}'_{i}
\end{bmatrix}$.

Consequently, the feasibility program can be formulated as follows
\begin{subequations}
\begin{align}
\label{32a} &\text{find}\quad \mathbf{U}\\
\label{32b}&\text{s.t.} \quad\,\, \eqref{17},\eqref{26a},\eqref{26b},\eqref{27},\eqref{28},\eqref{29},\eqref{30},\\
\label{32c}&\quad\quad\,\,\,\, \mathbf{U} \succeq \mathbf{0},\\
\label{32d}&\quad\quad\,\,\,\, \mathbf{U}_{m,m} \leq 1, 1\leq m\leq M, \mathbf{U}_{M+1,M+1}= 1,\\
\label{32e}&\quad\quad\,\,\,\, \text{rank}(\mathbf{U}) = 1.
\end{align}
\end{subequations}
However, problem \eqref{32a} is still non-convex due to the rank-one constraint \eqref{32e}. Similarly,  we use the SROCR method to tackle this problem. The relaxed rank-one constraint at $n$th iteration is given by
\begin{equation}
\label{33}
\mathbf{u}^{\text{eig-max},(n-1)}\mathbf{U}^{(n)}\mathbf{u}^{\text{eig-max},(n-1)}\geq u^{(n-1)}\text{Tr}(\mathbf{U}^{(n)}).
\end{equation}
Hence, we have the iterative convex feasibility program (ICFP) at $n$th iteration as
\begin{subequations}
\begin{align}
\label{34a} &\text{find}\quad \mathbf{U}\\
\label{34b}&\text{s.t.} \quad\,\, \eqref{32b}-\eqref{32d},\eqref{33}.
\end{align}
\end{subequations}
The iterative algorithm to solve problem \eqref{34a} is similar to \textbf{Algorithm-1} and is thus omitted for brevity.

The AO algorithm is summarized in \textbf{Algorithm-2}. The main computational complexity with the interior-point method is given by $\mathcal{O}\big(l_{\text{AO}}(l_{\text{a}}(3N_{\text{t}}^{2}+3)^{3.5}+l_{\text{p}}(M+1)^{7})\big)$, where the $l_{\text{a}}$ and $l_{\text{p}}$ denote the iteration numbers for solving ICP \eqref{25a} and ICFP \eqref{34a}, and $l_{\text{AO}}$ denotes the number of iterations.
\begin{table}[h!]
    \centering
    \begin{tabular}{p{235pt}}
    \toprule
    \textbf{Algorithm-2}: AO Algorithm for Solving Problem \eqref{11a} \\
    \midrule
    1: \textbf{Initialization}: Set $k = 1$, and initialize $\varpi^{(0)}$, $\tilde{\nu}^{(0)}$, $\mathbf{U}^{(0)}$, $w_{i}^{(0)}$, $\mathbf{w}_{i}^{\text{eig-max},(0)}$, $u^{(0)}$ and $\mathbf{u}^{\text{eig-max},(0)}$ ; \\
    2: \textbf{Repeat}: \\
    3: \quad Solve ICP \eqref{25a} with fixed $\mathbf{U}^{(k-1)}$;\\
    4: \quad Solve ICFP \eqref{34a} with fixed $\mathbf{W}_{i}^{(k-1)}$ and $\mathbf{W}_{\text{AN}}^{(k-1)}$;\\
    5: \quad $k = k+1$;\\
    6: \textbf{Until:} $|P_{\text{t}}^{(k)}-P_{\text{t}}^{(k-1)}|\leq \delta$.\\
    \bottomrule
    \end{tabular}
\end{table}

\section{Numerical Results}\label{4:Num-s}
\begin{figure}[t]
\centering
\includegraphics[scale=0.45]{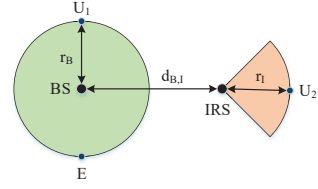}
\caption{Simulation setup with $\text{d}_{\text{B,I}}=50\text{m}$ and $\text{r}_{\text{B}}=\text{r}_{\text{I}}=2\text{m}$.}
\label{fig2:sim-model}
\end{figure}

In this section, we present numerical results to verify the performance of the proposed solution. The locations of all nodes are shown in Fig. \ref{fig2:sim-model}, and both the large and small scale fading are considered, i.e., $\mathbf{h}=d^{-\frac{\alpha}{2}}\mathbf{h}_{s}$, where $\mathbf{h}\in\{\mathbf{h}_{\text{I},i},\mathbf{h}_{\text{B},1},\mathbf{H}_{\text{B},\text{I}},
\mathbf{G}_{\text{B},\text{e}},\mathbf{G}_{\text{I},\text{e}}\}$. $d$ and $\alpha$ denote the distances and path-loss exponents, and $\mathbf{h}_{s}$ denotes the Rayleigh fading gain. Here, the path-loss exponents of the IRS-involved links (including BS-IRS link) are set as $2$, while the path-loss exponents of the BS-involved links are set as $4$. We adopt the normalized channel estimation error uncertainty for simulations, defined as $\xi_{\text{n}}=\frac{\varepsilon_{\text{e}}}{\|\mathbf{\hat{X}}\|_{F}}$. The convergence accuracy $\delta$ is set as $0.1$, and each point is averaged over $100$ trials.

\begin{figure}[t]
\centering
\includegraphics[scale=0.27]{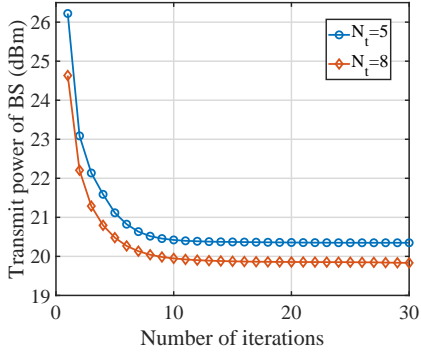}
\caption{Convergence of the proposed AO algorithm with $\xi_{\text{n}}=0.01$, $N_{\text{e}}=2$, $M=10$, $R_{\text{Q}}=1\text{bps/Hz}$, and $R_{\text{M}}=0.5\text{bps/Hz}$.}
\label{fig3:cvg}
\end{figure}

Fig. \ref{fig3:cvg} shows the convergence of the proposed AO algorithm with different $N_{\text{t}}$. It is observed from the figure that the transmit power monotonically decreases and converges very fast to a fixed value, which demonstrates the effectiveness of the proposed AO algorithm. It is also observed from the figure that an increase in $N_{\text{t}}$ can lower the transmit power but has no significant impact on the converge performance. That indicates properly increasing $N_{\text{t}}$ can benefit higher active beamforming gain for NOMA security enhancement without introducing large number of iterations for achieving convergence.

\begin{figure}[t]
\centering
\includegraphics[scale=0.27]{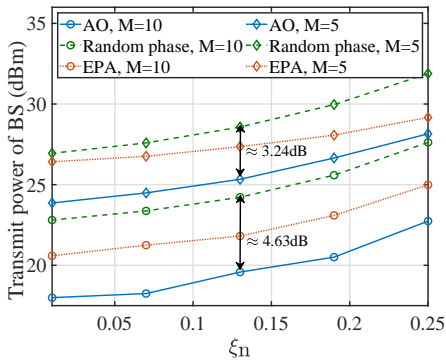}
\caption{Transmit power of BS versus the normalized channel estimation error with $N_{\text{t}}=8$, $N_{\text{e}}=2$, $R_{\text{Q}}=1\text{bps/Hz}$, and $R_{\text{M}}=0.5\text{bps/Hz}$.}
\label{fig4:Bas}
\end{figure}
\begin{figure}[t]
\centering
\includegraphics[scale=0.27]{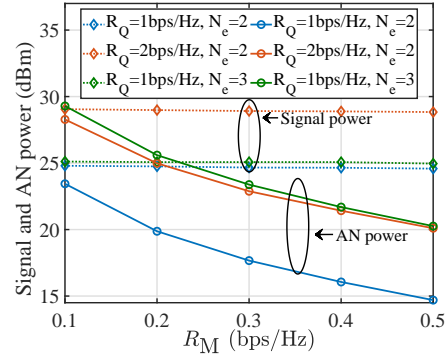}
\caption{Transmit power allocation between signals and AN versus the maximum eavesdropping rate at E with $N_{\text{t}}=8$, $M=5$, and $\xi_{\text{n}}=0.1$.}
\label{fig5:AN}
\end{figure}


Fig. \ref{fig4:Bas} compares the transmit power achieved by AO algorithm with two baseline schemes, i.e., random phase and equal power allocation (EPA) (i.e., $|\mathbf{w}_{1}|^{2}=|\mathbf{w}_{2}|^{2}$). As shown in Fig. \ref{fig4:Bas}, the total transmit power of three schemes monotonically decreases with the increased channel estimation error $\xi_{\text{n}}$, due to the fact that more transmit power is needed to compensate the increased channel uncertainty for secrecy guarantee. Compared with random phase and EPA, the proposed AO algorithm achieves the lowest transmit power consumption. Particularly, the random phase scheme has the worst performance and the performance gap between the random phase scheme and the proposed AO algorithm increases from $3.24\text{dB}$ to $4.63\text{dB}$ when  increasing $M$ from $5$ to $10$. This is because that the random phase shifts cannot always strengthen that at legitimate users and/or suppress the received signals at E. Furthermore, by increasing the reflecting elements, the passive beamforming gain can be improved, which is helpful to enhance the transmission security.


In Fig. \ref{fig5:AN}, the impact of the maximum eavesdropping rate on power allocation between signals and AN is plotted. As can be observed from Fig. \ref{fig5:AN}, the maximum eavesdropping rate and the number of antennas at E have significant impact on the AN power rather than the signal power. This can be understood as follows. An increase in the eavesdropping rate reduces the security requirement of network, thus less AN power is needed. While an increase in the number of antennas will strengthen the interception ability of E, and thus, the BS should allocate more power to AN to suppress eavesdropping. Whereas both the maximum eavesdropping rate and the number of antennas at E have little impact on signal transmission. Furthermore, the QoS requirement of users has a great effect on both signal and AN power. This is due to the fact that an increase in the QoS constraint results in a higher transmit power, which in turns improves the received signal strength of E. Hence, more power are needed for AN to degrade the reception quality of E.

\section{Conclusion}\label{5:Con}
This letter proposed a robust beamforming scheme to enhance secrecy of the IRS assisted NOMA network against a multi-antenna eavesdropper. An efficient AO algorithm was developed to optimize transmit beamforming and IRS reflection coefficients for transmit power minimization. Numerical results were provided to validate the security effectiveness of the proposed scheme and obtain valuable design insights into the robust design of secure transmission via IRS.



\begin{thebibliography}{99}

\bibitem{R.Zhang_IRS_magazine}
Q. Wu and R. Zhang, ``Towards smart and reconfigurable environment: Intelligent reflecting surface aided wireless network,''
\textit{IEEE Commun. Mag.}, vol. 58, no. 1, pp. 106--112, Jan. 2020.

\bibitem{C.Huang_TWC2019}
C. Huang, A. Zappone, G. C. Alexandropoulos, M. Debbah, and C. Yuen, ``Reconfigurable intelligent surfaces for energy efficiency in wireless communication,''
\textit{IEEE Trans. Wireless Commun.}, vol. 18, no.~8, pp. 4157--4170, Aug. 2019.

\bibitem{Q.Wu_TWC2019}
Q. Wu and R. Zhang, ``Intelligent reflecting surface enhanced wireless network via joint active and passive beamforming,''
\textit{IEEE Trans. Wireless Commun.}, vol. 18, no. 11, pp. 5394--5409, Nov. 2019.

\bibitem{L.Lv_magazine}
L. Lv, J. Chen, Q. Ni, Z. Ding, and H. Jiang, ``Cognitive non-orthogonal multiple access with cooperative relaying: A new wireless frontier for 5G spectrum sharing,''
\textit{IEEE Commun. Mag.}, vol. 56, no. 4, pp. 188--195, Apr. 2018.

\bibitem{L.Lv_NOMA_PLS}
L. Lv, H. Jiang, Z. Ding, L. Yang, and J. Chen, ``Secrecy-enhancing design for cooperative downlink and uplink NOMA with an untrusted relay,''
\textit{IEEE Trans. Commun.}, vol. 68, no. 3, pp. 1698--1715, Mar. 2020.

\bibitem{B.Zheng_IRS_NOMA}
B. Zheng, Q. Wu, and R. Zhang, ``Intelligent reflecting surface-assisted multiple access with user pairing: NOMA or OMA?''
\textit{IEEE Commun. Lett.}, vol. 24, no. 4, pp. 753--757, Apr. 2020.

\bibitem{X.Mu_IRS_NOMA}
X. Mu, Y. Liu, L. Guo, J. Lin, and N. Al-Dhahir, ``Exploiting intelligent reflecting surfaces in multi-antenna aided NOMA systems,''
\textit{IEEE Trans. Wireless Commun.}, doi: 10.1109/TWC.2020.3006915.

\bibitem{Z.Ding_IRS_NOMA_1}
Z. Ding and H. V. Poor, ``A simple design of IRS-NOMA transmission,''
\textit{IEEE Commun. Lett.}, vol. 24, no. 5, pp. 1119--1123, May. 2020.






\bibitem{X.Guan_IRS_AN}
X. Guan, Q. Wu, and R. Zhang, ``Intelligent reflecting surface assisted secrecy communication: Is artificial noise helpful or not?''
\textit{IEEE Wireless Commun. Lett.}, vol. 23, no. 9, pp. 1488--1492, Sept. 2019.

\bibitem{H.Shen_IRS_Security}
H. Shen, W. Xu, S. Gong, Z. He, and C. Zhao, ``Secrecy rate maximization for intelligent reflecting surface assisted multi-antenna communications,'' \textit{IEEE Commun. Lett.}, vol. 23, no. 9, pp. 1488--1492, Sep. 2019.

\bibitem{Z.Chu_IRS_Security}
Z. Chu, W. Hao, P. Xiao, and J. Shi, ``Intelligent reflecting surface aided multi-antenna secure transmission,''
\textit{IEEE Wireless Commun. Lett.}, vol.~9, no. 1, pp. 108--112, Jan. 2020.


\bibitem{X.Yu_imperfectCSI}
X. Yu, D. Xu, Y. Sun, D. W. K. Ng, and R. Schober, ``Robust and secure wireless communications via intelligent reflecting surfaces,''
\textit{IEEE J. Sel. Areas Commun.}, doi: 10.1109/JSAC.2020.3007043.

\bibitem{P.Cao_SROCR}
P. Cao, J. Thompson, and H. V. Poor, ``A sequential constraint relaxation algorithm for rank-one constrained problems,''
in \textit{Proc. 25th Eur. Signal Process. Conf. (EUSIPCO)}, Kosovo, 2017, pp. 1060--1064.

\bibitem{L.Lv_Secure_MISO}
L. Lv, Z. Ding, Q. Ni, and J. Chen, ``Secure MISO-NOMA transmission with artificial noise,''
\textit{IEEE Trans. Veh. Technol.}, vol. 67, no. 7, pp. 6700--6705, Jul. 2018.




\bibitem{Z.-Q.Luo_S-Procedure}
Z.-Q. Luo, J. F. Sturm, and S. Zhang, ``Multivariate nonnegative quadratic mappings,''
\textit{SIAM J. Optim.}, vol. 14, no. 4, pp. 1140--1162, Jul. 2004.




\end{thebibliography}
\end{document}